\documentclass{ws-ijmpcs}

\begin{document}

\markboth{S.~Bianco, D.~Pedrini, A.~Reis}
{Charm Hadronic Decays From FOCUS: Lessons Learnt}

%
\catchline{}{}{}{}{}
%

\title{CHARM HADRONIC DECAYS FROM FOCUS: LESSONS LEARNT}

\author{STEFANO BIANCO\footnote{Presenter}}

\address{Laboratori Nazionali di Frascati dell'INFN, v. E.~Fermi 40 \\
Frascati, 00044 Italy \\
stefano.bianco@lnf.infn.it}

\author{DANIELE PEDRINI}

\address{INFN Milano Bicocca, Edificio U2 - Piazza della Scienza 3 \\
I-20126 Milano - Italy\\
daniele.pedrini@mib.infn.it}

\author{ALBERTO REIS}

\address{Centro Brasileiro de Pesquisas F\'isicas,
Rua Dr. Xavier Sigaud, 150 - Urca \\
 Rio de Janeiro - RJ - Brasil 22290-180 \\
alberto@cbpf.br}

\maketitle

\begin{history}
\received{Day Month Year}
\revised{Day Month Year}
\end{history}

\begin{abstract}
The FOCUS photoproduction experiment took data in the ninenties and
produced a wealth of results in charm physics. Some of the studies
were seminal for contemporary experiments, and even paved the way for
the technology of many charm and beauty analysis tools.
\end{abstract}
\ccode{PACS numbers: 14.65.Dw; 14.40.Lb}
\section{Introduction}	
A retrospective discussion of selected FOCUS hadronic decays results
which have interest and implications for ongoing and future studies in
charm and beauty decays is presented. Topics which had long lasting
effects on heavy flavour physics are discussed, such as 
the measurement of
  $y_{CP}$ parameter in D mixing,  the study of   T-odd
correlations, and the four-body amplitude analysis. 
\par
 Data were collected 1996-1997 with the FOCUS spectrometer at the
 Fermilab Tevatron Wide Band photon beam.   The FOCUS
  spectrometer\cite{Frabetti:1990au}  was successor to E687 and 
    designed to
  study charm particles produced by 200 GeV photons using a fixed
  target spectrometer with vertexing, cerenkov detectors, em
  calorimeters, and muon id capabilities. Member groups of the FOCUS
  collaboration were from USA,
  Italy, Brazil, Mexico, Korea.    
\section{The $y_{CP}$ parameter}
\begin{figure}[h!]
\begin{center}
\includegraphics[height=2.in]{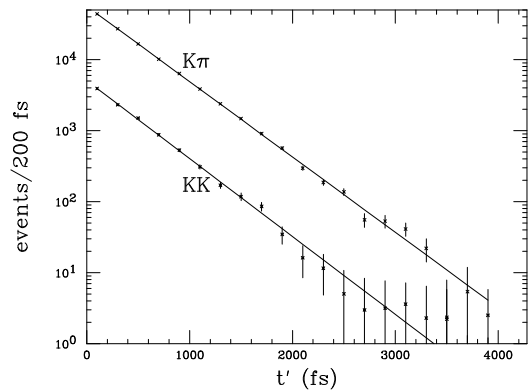}
\end{center}
\caption{ Signal versus reduced proper time for $D^0 \rightarrow
K^-\pi^+~ ~{\rm and}~~K^- K^+$ requiring $W_\pi - W_K > 4$
and $\ell/\sigma > 5$. The fit is over 20 bins of 200 fs bin width.
The data is background subtracted and includes the (very small)
Monte Carlo correction.  }
\label{life}
\end{figure}

The study \cite{Link:2000cu} 
consisted of a comparison of the lifetime of a CP even
final state, $D^0 \rightarrow K^- K^+$, to the lifetime of a CP
mixed decay, $D^0 \rightarrow K^- \pi^+$. The lifetime measurements
(Fig.~\ref{life}) 
were made using high signal-to-background $D^0$ samples consisting
of 10\,331 decays
into $K^- K^+$, and 119\,738 decays into $K^-
\pi^+$.  The measurement of lifetime differences 
corresponded to a direct  measurement of the mixing parameter $y_{\rm
CP} = ( \Gamma ({\rm CP~even}) - \Gamma ({\rm CP~odd}))/
(\Gamma({\rm CP~even}) + \Gamma ({\rm CP~odd})) =
0.0342 \pm 0.0139 \pm 0.0074 $.
\par
\begin{figure}
 \begin{center} 
\includegraphics[width=7cm]{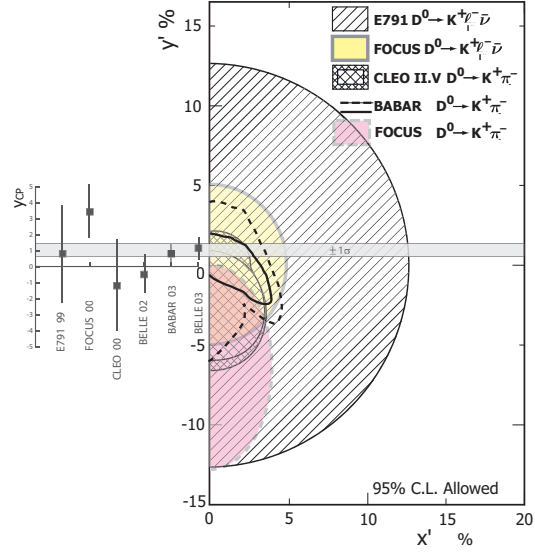} 
 \end{center}    
\caption{Summary of $x', y', y_{CP}$ measurements in 2003.}
 \label{FIG:YVSX} 
 \end{figure}

Such a first evidence for nonzero mixing was  followed by
higher-statistics
measurements from B-factories. Fig.\ref{FIG:YVSX} from  a 2003 review
paper \cite{Bianco:2003vb} shows 
the FOCUS measurement compared to the existing studies. With a
moderate significance due to low statistics, FOCUS showed evidence for
a nonzero lifetime difference.
Recent results from BABAR and BELLE have solidly established
the observation of D mixing.
Algorithms and selections used by FOCUS
may be used by forward-geometry experiment LHCb at the CERN LHC
 with much larger statistics very soon. 
\par
\section{T-odd correlations}
Triple-product correlations of the form
$\vec{v_1}\cdot(\vec{v_2}\times\vec{v_3})$, where each 
$\vec{v_i}$ is
a spin or momentum, are odd under time reversal (\emph{T}).
 By the
\emph{CPT} theorem, a nonzero value for these correlations would
also be a signal of \emph{CP} violation. A nonzero triple-product
correlation is evidenced by a nonzero value of the
asymmetry
\begin{equation}
  A_{T}  \equiv
\frac{\Gamma(\vec{v_1}\cdot(\vec{v_2}\times\vec{v_3})>0) -        
             \Gamma(\vec{v_1}\cdot(\vec{v_2}\times\vec{v_3})<0)}
{\Gamma(\vec{v_1}\cdot(\vec{v_2}\times\vec{v_3})>0) +             
        \Gamma(\vec{v_1}\cdot(\vec{v_2}\times\vec{v_3})<0)}  	
\end{equation}
where $\Gamma$ is the decay rate for
the process. There is a well-known technical complication:
	strong phases can produce a nonzero value of $A_{T}$, even if
	the weak phases are zero, that is \emph{CP} and \emph{T} 
	violation are not necessarily present. Thus, strictly
	speaking, the asymmetry $A_{T}$ is not in fact a
	\emph{T}-violating effect. Nevertheless, one can still
	obtain
a true \emph{T}-violating signal which takes care of the
complications due to strong phases by measuring a
	nonzero value of
\begin{equation}
A_\mathrm{Tviol} 
	\equiv \frac{1}{2}(A_{T}-\overline{A_{T}})
\end{equation}
where $\overline{A_{T}}$ is the
\emph{T}-odd asymmetry measured in the \emph{CP}-conjugate
decay process.
 This study was
inspired by a paper of Ikaros Bigi~\cite{Bigi}. In his paper
Bigi suggested a search for \emph{T} violation by looking
at the triple-product correlation (using the momenta of
the final state particles) in the decay mode \hbox{$D^0
\to K^-K^+\pi^-\pi^+$}. Such a correlation must necessarily
involve at least four final-state particles. This can be
understood by considering the rest frame of the decaying
particle and invoking momentum conservation. The number of
independent three-momenta is one less than the number of
final-state particles, so a triple product composed
entirely of momenta requires four particles in the final
state. We calculate $A_\mathrm{Tviol}$
for the decay modes $D^0 \to K^-K^+\pi^-\pi^+$ and
$D^+_{(s)} \to K^0_S K^+\pi^-\pi^+$ using data from the
FOCUS experiment, finding, with about 800 $D^0$ events \cite{Link:2005th},
${A_\mathrm{Tviol} (D^0) } = 0.010 \pm 0.057(\mathrm{stat.}) \pm
0.037(\mathrm{syst.})$, 
and,  with about 500 $D^+$ and 500 $D^+_s$ events,
${A_\mathrm{Tviol} (D^+) } = 0.023 \pm 0.062(\mathrm{stat.})
\pm 0.022(\mathrm{syst.})$,
 and
${A_\mathrm{Tviol} (D^+_s) } = -0.036 \pm 0.067(\mathrm{stat.}) \pm
0.023(\mathrm{syst.})$.   
\par
  In 2010 BABAR published  new limits\cite{delAmoSanchez:2010xj}
   by using the same
  FOCUS formalism. With $4.7\, 10^4$ events, BABAR measured
  ${A_\mathrm{Tviol} (D^+) } = 0.0010 \pm 0.0051(\mathrm{stat.})
\pm 0.0044(\mathrm{syst.})$.   Another order of magnitude limit reduction is
expected with the Frascati and KEK super-b factories data samples.
For an updated discussion see\cite{Pedrini:2010}.
\section{Four-body amplitude analysis}
FOCUS published the first
study\cite{Link:2007fi},\cite{Link:2004wx},\cite{Link:2003pt}  with a complete 
 4-body formalism of KK$\pi\pi$, KKK$\pi$, $\pi\pi\pi\pi$.
In this conference report only the $\pi\pi\pi\pi$ channel is discussed, with
6.3k events. This decay has a direct connection to decays used for
extraction of CKM parameters in B decays such as the angle $\alpha$
  from $B \rightarrow \rho^0 \rho^0$.
\par
A simple isobar model was used with a total of nine amplitudes and 16 free
parameters: $D\rightarrow \rho^0 \rho^0$; $D\rightarrow a_1\pi, a1\rightarrow
\rho^0\pi$; $D\rightarrow R \pi\pi,R\equiv (\sigma, \rho^0, f_0(980),
F_2(1270))$. As customary in all isobar models, final state interactions
(FSI) are not fully accounted for. Several
 helicity states were considered in
the fit functions. The study showed how the dominant
contribution to the $\pi\pi\pi\pi$ state comes from the $a_1^+(1260)\pi^-$
mode, which accounts  for over 60\% of the total decay rate.
The remaining part of the decay rate is equally divided between
$\rho^0 \rho^0$ and the quasi-three-body modes $D\rightarrow R\pi\pi$.
 The scalar component is dominated by the decay $D^0\rightarrow 
 \sigma \pi\pi$. The fit get significantly worse if  the 
 $\sigma \pi \pi$ amplitude is
replaced by an uniform non-resonant component. 
\par
Most important, although the one-dimensional projections suggest a
good fit, the goodness-of-fit tested on the full 5-dimensional phase
space is poor. This indicates that other effects need to be taken into
account for a more realistic 
model. One important ingredient is the final state interaction (FSI),
which could introduce energy-dependent phases. One needs to build a
model for the FSI based on chiral perturbation theory, but this is
rather challenging even for three-body final states. Another effect
that could play  a role  is the Bose-Einstein correlation between the
two pairs of identical pions. 
\par
 The final message is that, at
least in four-body decays, the strong interaction phases are still not
under control, which makes the extraction of weak phases governing CP
violation  rather uncertain. 
\section{Epilogue: on the importance of Dalitz analysis in the
extraction of CKM parameters }
Dalitz plot analysis is fundamental for a meaningful investigation of
decay modes in which resonances are present. As an example, FOCUS
showed\cite{Pedrini:2002xq} in the analysis of the decay mode 
$D^0\rightarrow K_sK^+K^-$
that a contamination from $f_0(980)$ under the $\phi(1020)$ signal is
always present if a simple mass cut is applied. This, however, changes
the assignment of the final state, since $K_s\phi$ is a CP-odd state,
while $K_sf_0$ is CP-even.
\par
These considerations forced, in early 2000s, the B-factory experiments
to revisit their analyses of some decay modes such as 
$B^0\rightarrow K_sK^-K^+$. Nowadays these analyses are performed with
the Dalitz plot technique.
\section{Conclusions}
FOCUS/E687 was a pioneering expedition towards the charm unexplored
territories, the first experiment which had important measurements
for all weakly decaying charm hadrons including weak lifetimes up to
the $\Omega_c$, thus showing us that some 
non-perturbative QCD effects could be understood. FOCUS gave hints of
unexpected phenomena such as a positive value for the $y_{CP}$
parameter in charm mixing, which is still today debated subject of
studies, and possibly the only mixing parameter significantly different
from zero observed so far. FOCUS did pioneering work for three-body
final states to help us in the future to find CPV in D decays, and to
help understand CP searches in B decays.Future searches for CPV from
New Physics need much more experimental and theoretical work. Finally,
FOCUS did pioneering work in $D \rightarrow 4 h$, useful for the
future and showing that more theoretical work is needed. From the
experimental side, while BaBar/BELLE reached the 1\% level in CPV
limits, much more statistics is needed to reach the 0.1\% level where
NP can generate effects.
\par
Three FOCUS hadronic
decays results have been singled out and discussed.
 The results have implications on upcoming
new results from 
the contemporary charm and beauty experiments. Algorithm and
selection used by FOCUS in the study of $y_{CP}$ parameter 
in D mixing may be used
by forward-geometry  experiment
 LHCb with much larger statistics. The formalism for 
 the search of CP violation via T-odd correlations was developed, and 
 first limit
on KK final state published, which was followed in 2010 by the 
new limit from BABAR in 2010 using the same formalism. 
 The first study with a complete 4-body formalism of
KK$\pi\pi$, KKK$\pi$,  $\pi\pi\pi\pi$ provided a  
direct connection to decays used for extraction of
CKM angle $\alpha$   in B decays: the poor 5D fit quality which emerges in
high-statistics studies implies the need for a
 better understanding of nonresonant
component. Finally, Dalitz plot analyses from FOCUS
showed the  importance of studying the resonant structure of hadronic
decays: S-waves  cannot be taken out with a simple mass cut. 
\section*{Acknowledgments}
The authors acknowledge valuable comments and discussions with I.I.~Bigi.
This research was supported in part by the Italian Istituto Nazionale
di Fisica Nucleare and Ministero della Istruzione Universit\`a e Ricerca,
.



\end{document}